\documentclass[prc,twocolumn]{revtex4}
\usepackage{amsmath,amssymb,epsfig,bm}
\newcommand{\be}{\begin{equation}}
\newcommand{\ee}{\end{equation}}
\newcommand{\bea}{\begin{eqnarray}}
\newcommand{\eea}{\end{eqnarray}}

\newcommand{\anu}{\bar\nu}

\newcommand{\ep}{\epsilon}

\newcommand{\vecq}{\bm q}
\newcommand{\vecp}{\bm p}

\newcommand{\sigmavec}{{\bm \sigma}}

\newcommand{\ie}{{ \it i.e., }}

\begin{document}
\title{Response functions of cold neutron matter: density fluctuations}
\author{Armen Sedrakian and Jochen Keller}
\address{Institute for Theoretical Physics,
J. W. Goethe University, D-60438 Frankfurt-Main, Germany}

\date{\today}

\begin{abstract}
We compute the finite temperature 
density response function of nonrelativistic
cold fermions  with an isotropic condensate.   
The pair-breaking contribution to the response function
evaluated in the limit of small 
three-momentum transfers $q$ within an effective theory which 
exploits series expansion in powers 
of small $q/p_F$, where $p_F$ is the Fermi momentum. The
leading order $O(q^2)$ contribution is universal and 
depends only on two fundamental scales,  the Fermi 
energy and the pairing gap. The particle-hole Landau
Fermi-liquid  interaction contributes first at 
the next-to-leading order  $O(q^4)$. The scattering contribution 
to the polarization tensor is nonperturbative (in the above sense) 
and is evaluated numerically. The spectral functions of 
density fluctuations are constructed and the relevance 
of the $q^2$ scaling for the pair breaking neutrino emission from neutron 
stars is discussed.
\end{abstract}

\maketitle
\section{Introduction}

The long-wavelength, low-energy dynamics of fermionic systems 
is determined by their response functions to soft perturbations, 
which are characterized by length scales that are large compared 
to the inverse Fermi wave vector and energies that are small 
compared to the Fermi energy.  At zero temperature 
the response functions of pair-correlated nuclear matter 
have been studied long ago by Larkin and Migdal~\cite{Migdal_book}. 
Recently, the response  functions of pair-correlated 
nuclear systems at nonzero temperature received attention  in the
context of neutral current neutrino emission via
pair breaking and formation in compact 
stars~\cite{Sedrakian:2006ys,Kolomeitsev:2008mc,Steiner:2008qz,Leinson:2009mq} 
and neutrino scattering in 
supernovae~\cite{Kundu:2004mz}. The evaluation of the response 
functions involves typically a resummation 
of infinite number of finite temperature ring diagrams. 
In the unpaired limit of normal Fermi liquid, this resummation scheme 
reduced to the familiar random-phase-approximation (RPA); for 
recent applications in nucleonic  and neutron
star matter, see Refs.~\cite{Olsson:2004ea,Lykasov:2005xh,Margueron:2005nc,Baldo:2008pb}.

In attractive, cold, fermionic 
systems the gap $\Delta$ in the quasiparticle spectrum is small
compared to the Fermi energy and the hierarchy of energy
scales depends on the magnitude of the perturbation,
which can take arbitrary values with respect to the pairing gap 
$\Delta$. In this work we focus on density perturbations and show
that the two distinct contributions to the response function 
through the scattering and pair breaking processes are effective
below and  above the energy threshold 
$2\Delta$,  \ie the energy needed to break a pair. 
The pair breaking processes are of special importance for
applications in compact stars; to evaluate them, we propose a new systematic 
low-transferred-momentum expansion of the response function  
which builds on the previous work on polarization tensors 
of cold superfluid fermionic systems~\cite{Sedrakian:2006ys}. 
Specifically, we show that the 
pair breaking contribution to the polarization tensor posses 
a well-defined expansion with respect to the ratio of the 
momentum of the external current to the Fermi momentum of the 
fermions. We also adopt more general ansatz for the driving 
terms in the integral equations of  Ref.~\cite{Sedrakian:2006ys}  
by lifting the degeneracy among the particle-particle and 
particle-hole channels. We work in the nonrelativistic limit, \ie
the ratio of the Fermi  velocity to the speed of 
light is small  ($v_F/c\ll 1$).

The density response functions can be utilized to  
determine the spectrum of the collective modes,  the stability 
of the system toward clustering, the rates of electromagnetic 
and weak radiation processes, and so on.
In particular, the rates of neutrino reactions in stellar 
interiors can be expressed through the response function of 
underlying matter to vector and axial-vector weak 
currents~\cite{Burrows:1998ek,Sedrakian:1999jh,Sedrakian:2000kc,Sedrakian:2006mq}. 
In nonrelativistic limit, the vector and axial-vector 
responses are mapped onto the responses to the density 
and spin-density perturbations, respectively. 
The response functions in the superfluid neutron matter 
were recently computed and the neutrino emission 
rates were determined in Refs.~\cite{Sedrakian:2006ys,Kolomeitsev:2008mc,Steiner:2008qz,Leinson:2009mq}. 
Phenomenologically, these are important in modeling the 
cooling of intermediate age neutron stars and the 
superburst in accreting neutron 
stars~\cite{Brown:2009kw,Page:2009fu,Gupta:2006fd}. 

It is now well established that at zeroth order ($\vecq=0$) the 
pair breaking density response function vanishes, as required by the 
$f$-sum rule for the polarization tensor, which is a direct 
consequence of the baryon number conservation. 
Some authors found analytically the 
leading-order contribution to the polarization which 
arises at order $q^4$~\cite{Leinson:2006gf,Kolomeitsev:2008mc}. 
Since there is no general argument that requires the 
coefficient of the leading-order  $q^2$ term to be zero, 
the answer may depend on the approximations involved in the 
theory. In section~\ref{sec:small_mom_exp}
we study in detail a new small $q$ expansion of 
the polarization tensor obtained in Ref.~\cite{Sedrakian:2006ys}
and find that the coefficient of the 
$q^2$ term in the series expansion of the density response 
function is indeed nonzero. Furthermore, it turns out to be 
universal, \ie it depends only on two fundamental scales of the 
problem, the Fermi energy and pairing gap, and is independent 
of the strength of particle-hole interaction.

This paper is organized as follows. In Sec.~\ref{sec:density_response}
we derive the vertex functions and polarization tensor in a more general 
setting that in Ref.~\cite{Sedrakian:2006ys} by using different 
particle-particle and particle-hole interactions and clarify the 
approximations that arise in the weak coupling Bardeen-Cooper-Schrieffer
(BCS) limit. A small 
momentum transfer expansion is applied to the pair breaking 
polarization tensor in Sec.~\ref{sec:small_mom_exp}. We also show in 
Sec.~\ref{sec:scatter} the results of exact numerical evaluation 
of the scattering part of the polarization tensor.
Further in Sec.~\ref{sec:unpaired}  we verify that the unpaired and 
uncorrelated limits are recovered from the scattering part.
Our conclusions are collected in 
Sec.~\ref{sec:conclude}. Details of calculations are presented 
in  Appendices \ref{appendix_A} and \ref{appendix_B}. We use the
natural units $\hbar = c = 1$ throughout and assume that the
Boltzmann constant $k_B=1$.

\section{Density response function}
\label{sec:density_response}

In this section we derive the general form of the 
density-response function of 
neutron matter at nonzero temperature. At densities 
below the saturation  density neutron matter forms 
a $^1S_0$ pair condensate and can be described by the 
weak coupling limit of the BCS theory. This affects not only 
the approximations that are applied to the gap equation, 
but also the approximate relations between the loop integrals, 
as we discuss below.

The couplings in the particle-particle ($pp$) and particle-hole 
($ph$) channels, $v_{pp}$ and $v_{ph}$, are assumed zero range; 
often their values are taken to be degenerate $v_{pp} = v_{ph}$ 
and equal to the lowest order Landau parameter $f_0$. We shall 
lift this approximation below by assuming $v_{pp} \neq v_{ph}$.
The spectrum of paired neutrons is given by
\be
\epsilon_p = \sqrt{\xi_p^2 +\Delta^2(\vecp) },
\ee
where $\xi_p = p^2/2m^*  -\mu$ is the quasiparticle spectrum in the normal
state, $\Delta(\vecp)$ is the energy gap, $\vecp$ is the three-momentum,
 $m^*$ is the effective mass and $\mu$ is the chemical potential.
For contact pairing interaction the gap is momentum independent,
$\Delta(\vecp) \equiv \Delta$.

The softness of the modes implies that their wave vector
$\vert\vecq\vert\ll p_F$. Accordingly, we  write
\be
\xi_{p+q} 
 = \frac{p^2}{2m^*}
\left(1 + \frac{2\vecp\cdot\vecq}{p^2}+ \frac{q^2}{p^2}\right)-\mu ,
\ee
and consider the  second and third terms in the bracket as small 
compared to unity, since  $p\simeq p_F$, where $p_F$ is the 
Fermi momentum. Thus, we may write 
$\xi_{p+q} \simeq\xi_p +  \mu_0(2 y x+y^2)$, where
$ x = (\vecp\cdot\vecq)/(\vert p\vert\vert q\vert)$
 and $y = {q}/{p_F}$, whereby $y\ll 1$.
Here $\mu_0 = p_F^2/2m^*$ is the chemical potential
at zero temperature; we shall drop the 0 index in the following.
Several observations are in order:
\begin{enumerate}
\item If the expansion is carried out with respect to
the small parameter $\delta\xi = \xi_{p+q}-\xi_p$, as in
Ref.~\cite{Sedrakian:2006ys}, the power counting
is not manifest.
At the leading-order the terms which scale linearly in $x$ 
drop on angle integration in symmetrical limits.
The only nonzero contribution
proportional to $q^2$ is then furnished by the recoil term.
One needs to carry out the  $\delta\xi$ expansion at least to 
second order to obtain all relevant terms that are 
of order $q^2$.

\item An alternative to the expansion with respect to 
$\delta\xi$ is the
expansion with respect to the ratio $y = \vert \vecq\vert/p_F$.
This expansion exploits the softness of the modes and
applies both for timelike and
spacelike momentum transfers. It is an alternative 
to expansions in the ratios $\omega/qv_F$ and 
$qv_F/\omega$, which are valid in these regimes, respectively.

\item Organizing the expansion in powers of $v_F/c \ll 1$
(nonrelativistic fermions) does not guarantee {\it per se }
the convergence of the series. At any fixed density, the Fermi 
velocity $v_F$ is constant and for sufficiently large momentum transfers
($q \ge p_F/2$) the series will fail to converge.

\item Finally, the smallness of the expansion parameter
is necessary but not sufficient condition for the convergence of the
Taylor series. The validity of the expansion should be
checked by an exact numerical computation of the
loop integrals. 
\end{enumerate}
In the following we shall demonstrate that the pair breaking 
part of the response function can be expanded systematically 
with respect to the  $y$ parameter. Such expansion is thus 
valid for small three-momentum transfers, but arbitrary 
energy transfers. 


\subsection{Vertex functions}
We start with integral equations for the vertex functions and 
derive a (slight) generalization of their counterparts in 
Ref.~\cite{Sedrakian:2006ys} that distinguish the particle-particle 
and particle-hole interactions. These we write as sums of central 
and spin-spin interaction terms
\bea \label{LP}
V^{pp} &\simeq & v_{pp} + v_{pp} (\sigmavec \cdot \sigmavec')+\dots,\\
V^{ph} &\simeq & v_{ph} + v_{ph} (\sigmavec \cdot \sigmavec')+\dots,
\eea
where the ellipses stand for the tensor and spin-orbit terms that are 
subdominant at relevant densities in neutron matter. 

The integral equations defining the scalar vertex, which 
we write in an operator form, are given by~\cite{Migdal_book,Sedrakian:2006ys}
\bea\label{g1}
\hat\Gamma_1 &=& \Gamma_0 
             +  v_{ph} (G\Gamma_1G
             +       \hat F\hat\Gamma_3G
             +       G\hat\Gamma_2\hat F
             +       \hat F\Gamma_4\hat F),  \\ 
\label{g2}
\hat\Gamma_2 &=&  \quad \quad v_{pp} (G\hat\Gamma_2G^{\dagger}
             +       \hat F\Gamma_4G^{\dagger}
             +       G\Gamma_1\hat F
             +       \hat F\hat\Gamma_3\hat F), \\ 
\label{g3}
\hat\Gamma_3 &=&  \quad \quad v_{pp} (G^{\dagger}\hat\Gamma_3G
             +       \hat F\Gamma_1G
             +       G^{\dagger}\Gamma_4\hat F
             +       \hat F\hat\Gamma_2\hat F),\\
\label{g4}
\hat\Gamma_4 &=& \Gamma_0 
             +  v_{ph} (G^{\dagger}\Gamma_4G^{\dagger}
             +       \hat F\Gamma_1 \hat F
             +      \hat F \hat\Gamma_2 G^{\dagger}
             +       G^{\dagger}\hat\Gamma_3\hat F).  \nonumber\\
\eea
Here $\hat F = -i\sigma_y F$,  $\sigma_y$ is the second component 
of the Pauli matrix,  $\Gamma_0 = 1$. When $v_{pp}=v_{ph}$, 
Eqs.~(\ref{g1})-(\ref{g4}) become identical to those of 
Ref.~\cite{Sedrakian:2006ys}. Let us now define the ``elementary loop''
as 
\be 
\Pi_{XX'}(q) = g\int\frac{d^4p}{(2\pi)^4} X(p)X'(p+q),
\ee 
where $X\in \{G,G^{\dagger},F,F^{\dagger}\}$ and $g$ is the degeneracy 
factor, which we omit in the intermediate equations and restore in the 
final ones. Direct calculations show that 
\be
\Pi_{G^{\dagger}F} = \Pi_{FG},\quad \Pi_{GF} = \Pi_{FG^{\dagger}},
\quad \Pi_{G^{\dagger}G^{\dagger}} = \Pi_{GG};
\ee
these equalities imply that $\Gamma_1=\Gamma_4$, which is 
a  consequence of the time-reversal invariance of the system. 
\begin{widetext}
The remaining equations read
\be\label{G1}
\left(\begin{array}{ccc}
1- v_{ph}[\Pi_{GG}-\Pi_{FF}] &v_{ph}\Pi_{GF} & 
v_{ph}\Pi_{FG}\\
-2v_{pp}\Pi_{GF} & [1-v_{pp}\Pi_{GG^{\dagger}}] &v_{pp} \Pi_{FF}\\
-2v_{pp}\Pi_{FG} & v_{pp}\Pi_{FF} & [1-v_{pp}\Pi_{G^{\dagger} G}]
\end{array}\right)\left(\begin{array}{c}
\Gamma_1\\
\Gamma_2\\
\Gamma_3       
\end{array}\right) = \left(\begin{array}{c}
\Gamma_0\\
0\\
0 
\end{array}\right).
\ee
\end{widetext}
There are six distinct loops in Eq.~(\ref{G1}), namely
$\Pi_{GG}$, $\Pi_{FF}$, $\Pi_{GF}$, 
$\Pi_{FG}$, $\Pi_{G^{\dagger} G}$, $\Pi_{GG^{\dagger}}$. 
In the weak coupling limit 
$\Pi_{GF} \simeq -\Pi_{FG}$, $\Pi_{ G^{\dagger}G} \simeq \Pi_{GG^{\dagger}}$, 
an approximation discussed in detail in Sec.~\ref{sec:pol_function}.
This reduces the number of equations from three to two
\bea
\left(\begin{array}{ccc}
1- v_{ph}{\cal A} &v_{ph}{\cal B} \\
-v_{pp}{\cal B} & -v_{pp} {\cal C}\\
\end{array}\right)\left(\begin{array}{c}
\Gamma_1\\
\Gamma_2\\
\end{array}\right) = \left(\begin{array}{c}
\Gamma_0\\
0\end{array}\right),
\eea
where 
\bea \label{A_loop}
{\cal A}(q) &=& \Pi_{GG}(q)-\Pi_{FF}(q),\\
\label{B_loop}
{\cal B}(q) &=& 2\Pi_{FG}(q),\\
\label{C_loop}
{\cal C}(q) &=&\Pi_{GG^{\dagger}}(q) + \Pi_{FF}(q)-(v_{pp})^{-1},
\eea
with $q = (\omega, \vecq)$. The solutions for the remaining 
two vertex functions reads
\bea\label{Gx1} 
\Gamma_1(q) &=& 
\frac{\Gamma_0 {\cal C}(q)}{{\cal C}(q)- v_{ph}
[{\cal A}(q) {\cal C}(q) + {\cal B}(q)^2]},
\\ 
\label{Gx2} 
\Gamma_2(q) &=& -\frac{\Gamma_0 {\cal B}(q)}{{\cal C}(q)- v_{ph}
[{\cal A}(q) {\cal C}(q) +{\cal B}(q)^2]}.
\eea
When $v_{pp}=v_{ph}$ these reduce to Eqs. (\ref{Gx1}) 
and  (\ref{Gx2})  of 
Ref.~\cite{Sedrakian:2006ys}. 
It seen that the interaction in the $pp$ channel is 
absorbed in the gap equation and it is the $ph$ interaction that 
enters the renormalization of the one-loop polarization tensor. 
This result could have been  anticipated from the limiting form of 
the RPA polarization tensor of normal Fermi liquids (see 
Sec.~\ref{sec:unpaired}).

\subsection{Polarization tensor}
\label{sec:pol_function}
The full polarization tensor is given by  Eq. (35) of 
Ref.~\cite{Sedrakian:2006ys} with the replacement $v\to v_{ph}$ 
\be \label{FULL_PI}
\Pi^R(q) = \frac{{\cal A}(q){\cal C}(q)+{\cal B}(q)^2}
{{\cal C}(q) -v_{ph}[{\cal A}(q){\cal C}(q)+{\cal B}(q)^2]}.
\ee
\begin{widetext}
The ``elementary loops'' are defined explicitly as
\bea
\label{A1}
\Pi_{GG}(q)
&=& \int\frac{d^3p}{(2\pi)^3} \Biggl\{
     \left[ \frac{u_p^2u_k^2}{iq+\ep_p-\ep_k}-
             \frac{v_p^2v_k^2}{iq-\ep_p+\ep_k}
     \right]\left[f(\ep_p)-f(\ep_k)\right]\nonumber\\
&+&  \left[ \frac{u_k^2v_p^2}{iq-\ep_p-\ep_k}-
             \frac{u_p^2v_k^2}{iq+\ep_p+\ep_k}
     \right]\left[f(-\ep_p)-f(\ep_k)\right]\Biggr\},\\
\label{A2bis}
\Pi_{FG}(q)
&=& - \int\frac{d^3p}{(2\pi)^3}  \Biggl\{
u_pv_p \left[ \frac{u_k^2}{iq+\ep_p-\ep_k}+
             \frac{v_k^2}{iq-\ep_p+\ep_k}
     \right]\left[f(\ep_p)-f(\ep_k)\right]\nonumber\\
&-& u_pv_p \left[ \frac{u_k^2}{iq-\ep_p-\ep_k}+
             \frac{v_k^2}{iq+\ep_p+\ep_k}
     \right]\left[f(-\ep_p)-f(\ep_k)\right]\Biggr\},\\
\label{A4}
\Pi_{FF}(q)
&=& \int\frac{d^3 p}{(2\pi)^3}  \Biggl\{
 u_pu_kv_pv_k \Biggl\{\left[\frac{1}{iq+\ep_p-\ep_k}-
             \frac{1}{iq-\ep_p+\ep_k}
     \right]\left[f(\ep_p)-f(\ep_k)\right]\nonumber\\
&+&  \left[ \frac{1}{iq+\ep_p+\ep_k}-
             \frac{1}{iq-\ep_p-\ep_k}
     \right]\left[f(-\ep_p)-f(\ep_k) \right]\Biggr\}\Biggr\},\\
\label{A5}
\Pi_{G^{\dagger}G}(q)
&=& - \int\frac{d^3p}{(2\pi)^3}  \Biggl\{
 \left[ \frac{u_k^2v_p^2}{iq+\ep_p-\ep_k} -
             \frac{u_p^2v_k^2}{iq-\ep_p+\ep_k}
     \right]\left[f(\ep_p)-f(\ep_k)\right]\nonumber\\
&+&  \left[ \frac{u_p^2u_k^2}{iq-\ep_p-\ep_k}-
             \frac{v_p^2v_k^2}{iq+\ep_p+\ep_k}
     \right]\left[f(-\ep_p)-f(\ep_k)\right]\Biggr\},
\eea
\end{widetext}
where $k = p + q$ and the coherence factors are given by
$u_p^2 = (1/2)[1+  \xi_p/\ep_p]$ and $v_p^2 = 1-u_p^2$.

The expression for $\Pi_{FG}(q)$ above applies at arbitrary couplings;
however Eq.~(\ref{B_loop}) presumes weak-coupling approximation, because
$\Pi_{GF}(q) = - \Pi_{FG}(q)$ holds only in this limit. The weak-coupling
limit for the function $\Pi_{FG}(q)$ is obtained on substituting in square
braces $u_k^2 = u_p^2 = v_p^2 = v_k^2 = 1/2$. This statement is equivalent
to ignoring integrals of the type
\bea \label{weak_int}
I &=& \int\frac{d^3p}{(2\pi)^3} \frac{\Delta^2}{2\ep_p}\frac{\xi_p}{\ep_p}
\frac{f(\ep_p)-f(\ep_k)}{iq+\ep_p-\ep_k}\simeq 0 .
\eea
The integral vanishes because $\xi_p$ changes sign for momenta
above and below the Fermi momentum, while the remainder of the
integrand is an even function in the vicinity of $p_F$. Thus
the loop polarization function reduces to
\bea
\label{A2_1}
\Pi_{FG}(q)
&=& - \frac{1}{2}\int\frac{d^3p}{(2\pi)^3}  \Biggl\{
u_pv_p \nonumber\\
&&\left[ \frac{1}{iq+\ep_p-\ep_k}+
             \frac{1}{iq-\ep_p+\ep_k}
     \right]\left[f(\ep_p)-f(\ep_k)\right]\nonumber\\
&-& u_pv_p \left[ \frac{1}{iq-\ep_p-\ep_k}+
             \frac{1}{iq+\ep_p+\ep_k}
     \right]\nonumber\\
&&\left[f(-\ep_p)-f(\ep_k)\right]\Biggr\}.
\eea
The form of the polarization loop $\Pi_{G^{\dagger}G}(q)$ is valid for arbitrary
couplings. The relation $\Pi_{G^{\dagger}G}(q) = \Pi_{GG^{\dagger}}(q)$ is
established on noting that, for example, $u_p^2v_k^2 =
(u_p^2v_k^2 - u_k^2v_p^2)+u_k^2v_p^2$ and that the combination in braces
vanishes in the weak coupling as it leads to an integral 
of the type (\ref{weak_int}).

The contributions to the polarization function due to
quasiparticle scattering and pair breaking separate. For the
contributions from  scattering the poles are located
at $\pm(\ep_p -\ep_k)$ and the distribution is given by
the combination $f(\ep_p)-f(\ep_k)$. For the
pair breaking contributions the poles are located at $\pm(\ep_p+\ep_k)$
and the distribution is proportional $1-f(\ep_p)-f(\ep_k)$. The
pair breaking contribution vanishes in the limit $T\to T_c^{-}$ ($T$
is the temperature, $T_c$ is the critical temperature of
the phase transition).

Now we write the retarded polarization functions $\Pi^R(q)$ in terms of
the functions ${\cal A}(q)$, ${\cal B}(q)$ and ${\cal C}(q)$
after performing analytical continuation ($iq \to \omega + i\delta$)
in functions $\Pi_{GG}(q)$, $\Pi_{FF}(q)$,  $\Pi_{FG}(q)$
$\Pi_{G^{\dagger}G}(q)$. After some straightforward
algebraic transformations we find
\begin{widetext}
\bea\label{A2_3}
{\cal A}(q)  &=&  \int\frac{d^3p}{(2\pi)^3}
\Biggl\{\frac{1}{2}(\ep_k-\ep_p)
\left[1+\frac{\xi_p\xi_k}{\ep_p\ep_k}-\frac{\Delta^2}{\ep_p\ep_k}\right]
\frac{f(\ep_p)-f(\ep_k)}
{(\omega+i\delta)^2-(\ep_p-\ep_k)^2}\nonumber\\
&+&\frac{1}{2}(\ep_p+\ep_k)
\left[1-\frac{\xi_p\xi_k}{\ep_p\ep_k}+ \frac{\Delta^2}{\ep_p\ep_k}
\right]\frac{f(-\ep_p)-f(\ep_k)}
{(\omega+i\delta)^2-(\ep_p+\ep_k)^2}\Biggr\},\\
\label{B2}
{\cal B}(q) &=&-\Delta\omega \int\frac{d^3p}{(2\pi)^3}
\frac{1}{\ep_p} \Biggl[\frac{f(\ep_p)-f(\ep_k)}{(\omega+i\delta)^2-(\ep_p-\ep_k)^2}
-\frac{f(-\ep_p)-f(\ep_k)}{(\omega+i\delta)^2-(\ep_p+\ep_k)^2}\Biggr],\\
\label{C2_3}
{\cal C}(q) &=& \frac{1}{2}
\int\frac{d^3p}{(2\pi)^3} \Biggl\{
\left[
 (\ep_p-\ep_k)\left(1-\frac{\xi_p\xi_k}{\ep_p\ep_k}
-\frac{\Delta^2}{\ep_p\ep_k}\right)\right]
\frac{f(\ep_p)-f(\ep_k)}{(\omega+i\delta)^2-(\ep_p-\ep_k)^2} \nonumber\\
&-&
\left[(\ep_p+\ep_k)\left(1+\frac{\xi_p\xi_k}{\ep_p\ep_k}
+\frac{\Delta^2}{\ep_p\ep_k}\right)\right]
\frac{f(-\ep_p)-f(\ep_k)}{(\omega+i\delta)^2-(\ep_p+\ep_k)^2}
 -  \frac{1}{\ep_p}\left[1-2f(\ep_p)\right]\Biggr\}.
\eea
\end{widetext}
Note that  $\lim_{\vecq\to 0,\omega\to 0}{\cal C}(q)= 0$,
since the coupling constant in the particle-particle channel can
be expressed as
\bea
(v_{pp})^{-1} &=& 
\Pi_{G^{\dagger}G}(q=0)  + \Pi_{FF}(q=0)\nonumber\\
&=& \int\frac{d^3p}{(2\pi)^3}
\frac{1}{2\ep_p}\left[1-2f(\ep_p)\right],
\eea
which is the gap equation for the contact interaction $v_{pp}$. We do not
need to specify the regularization of the gap equation, since its divergence 
is eliminated in the loop integrals. For numerical purposes we 
will adopt gaps obtained from finite-range interactions in 
Ref.~\cite{Sedrakian:2003cc}.

\section{Evaluating response functions}

Equations (\ref{A2_3})--(\ref{C2_3}) separate into scattering 
and pair breaking contributions. We shall see that the first 
contributes essentially below the pair breaking threshold 
$\omega<2\Delta$, whereas the second contributes for $\omega>2\Delta$.
In the following we discuss in detail the pair breaking part and
its small momentum expansion. The scattering part will be addressed 
later in this section, where we evaluate it numerically.

\subsection{pair breaking response function: small momentum expansion}
\label{sec:small_mom_exp}

The small-$y$ expansion of the pair breaking part of the 
polarization tensor is obtained upon writing
${\Pi}^R(q)= \sum_n {\cal P}_n y^n$,  ${\cal A}(q)= 
\sum_n {\cal A}_n y^n$ and similarly for the functions 
${\cal B}$ and ${\cal C}$ and truncating the Taylor series
at the desired order in $y$. The odd powers of $y$
do not contribute to the series.

Up to order $y^4$ the
coefficients for the polarization tensor are given by
\bea
{\cal P}_0 &=& 0,\\
{\cal P}_2 &=& \frac{ 2\,{\cal B}_0\,{\cal B}_2 +
       {\cal A}_2\,{\cal C}_0+ {\cal A}_0\,{\cal C}_2 }{{\cal C}_0},\\
{\cal P}_4 &=&  \frac{{\cal B}_2^2 + 2\,{\cal B}_0\,{\cal B}_4 +
            {\cal A}_4\,{\cal C}_0 + {\cal A}_2\,{\cal C}_2
	 +  {\cal A}_0\,{\cal C}_4}{{\cal C}_0}\nonumber\\
	 &-&  {\cal P}_2 \left[\frac{{\cal C}_2}{{\cal C}_0}
              -v_{ph}{\cal P}_2\right].\eea
The coefficients of the expansion are given by (for details see 
Appendix~\ref{appendix_A})
\bea
{\cal A}_0 & = & 2\Delta^2\int\!\!\frac{d^3p}{(2\pi)^3} \frac{1}{\ep_p}L_0,\\
{\cal B}_0 & = & \Delta\omega \int\!\!\frac{d^3p}{(2\pi)^3}\frac{1}{\ep_p} L_0,\\
{\cal C}_0 &=& -\frac{\omega^2}{2}\int\!\!\frac{d^3p}{(2\pi)^3} \frac{1}{\ep_p}L_0,
\eea
\bea
{\cal A}_2 & = & \int\!\!\frac{d^3p}{(2\pi)^3} \Biggl[
\frac{\Delta^2\,\mu\,\xi_p}{\ep_p^5}(6\mu\xi_px^2 -\ep_p^2 )L_0
\nonumber\\
&-&\frac{2\Delta^2\,\mu\,\xi_p}{\ep_p^3}~L_1 ~x^2~
+\frac{2\Delta^2}{\ep_p}L_2
\Biggr],\\
{\cal B}_2 & = & \Delta\omega \int\!\!
\frac{d^3p}{(2\pi)^3}\frac{1}{\ep_p} L_2 ,\\
{\cal C}_2 &=& -\int\!\!\frac{d^3p}{(2\pi)^3}\left[ \frac{\mu\,\xi_p}{\ep_p}~L_0
+\frac{2\mu\xi_p}{\ep_p}L_1 ~x^2 + 2\ep_p L_2 \right],\nonumber\\
\eea
where the functions $L_0$, $L_1$, and $L_2$ are defined 
by Eqs. (\ref{L0})--(\ref{L2}) of Appendix~\ref{appendix_A}.
The  zeroth order term ${\cal P}_0$ 
vanishes as a consequence of the $f$-sum rule. 
The leading order nonzero term is given by 
\bea
{\cal P}_2 = \Delta^2\nu(p_F) (I_0+I_1+I_2),
\eea
where $\nu(p_F) = gm^*p_F/2\pi^2$ is the density of states at
the Fermi surface, and
\bea\label{int_I0}
I_0 &=& 4\nu(p_F)^{-1}\mu\,\int\frac{d^3p}{(2\pi)^3}
\frac{\xi_p}{\ep_p\omega^2}\nonumber\\
&&\Bigl[1 +\frac{\omega^2}{4\ep_p^2}
\left(\frac{6\mu\xi_p}{\ep_p^2}x^2 -1\right)\Bigr]L_0,\\
\label{int_I1}
I_1 &=& 2\nu(p_F)^{-1}\mu\int\frac{d^3p}{(2\pi)^3}
\frac{\xi_p}{\ep_p^3}\left[\frac{4\ep_p^2}{\omega^2}-1\right]L_1 ~x^2,\\
\label{int_I2}
I_2 &=& 2\nu(p_F)^{-1}\int\frac{d^3p}{(2\pi)^3}
 \frac{1}{\ep_p}\left[\frac{4\ep_p^2}{\omega^2}-1\right]L_2.
\eea
These integrals are evaluated in Appendix~\ref{appendix_B}. 
We obtain the following
analytical result for the imaginary part of ${\cal P}_2$:
\bea\label{calP_final}
{\rm Im}{\cal P}_2
&=&-\frac{\pi\nu(p_F)}{3T}
   \frac{\Delta^2\mu\, {\rm sgn (\omega)}}{\omega^5
\sqrt{\omega^2-4\Delta^2}}\,{\rm sech}^2\left(\frac{\omega}{4T}\right)
\nonumber\\
&&\Bigg\{4\mu\omega\,(\omega^2-4\Delta^2) -
T\,[40\Delta^2\mu\nonumber\\
&-&3\omega^2\,(6\mu+\sqrt{\omega^2-4\Delta^2})]
  \sinh\left(\frac{\omega}{2T}\right)
  \Bigg\}\,\theta(\vert\omega\vert-2\Delta).\nonumber\\
\eea
To leading-order the imaginary part of the 
polarization tensor is then given by 
\be 
{\rm Im}{\Pi}^R(q) = {\rm Im}{\cal P}_2(\omega) 
\left(\frac{\vecq}{p_F}\right)^2+O(q^4).
\ee
\begin{figure}[t]
\begin{center}
\includegraphics[width=\linewidth,height=7.0cm,angle=0]{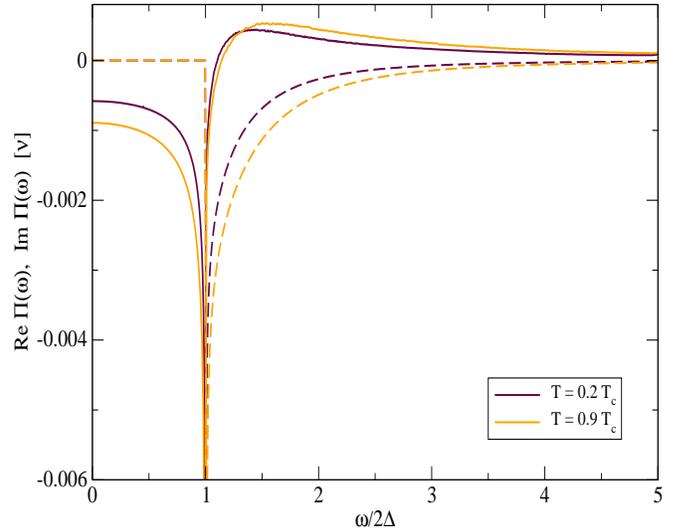}
\end{center}
\caption[]
{Dependence of the real (solid lines) and imaginary (dashed lines)  
parts of the pair breaking 
polarization tensor in units of density of states on 
the energy transfer in units of threshold energy $2\Delta(T)$
for fixed  momentum transfer $q=0.1 p_F$, with $p_F = 0.1$ 
fm$^{-1}$,  and two temperatures 
 $T = 0.2 T_c$ (heavy lines) and $T = 0.9T_c$ (light lines).
The zero temperature  gap is taken to be $\Delta (0) = 1.$ MeV, 
$T_c = \Delta(0)/1.76$. 
}\label{fig:reimpi}       
\end{figure}
\noindent
The real part of the polarization tensor 
follows from the dispersion (Kramers-Kronig) relation:
\be\label{KK_relation} 
{\rm Re}\Pi^R (\omega,\vecq)= 
-\frac{1}{\pi}\int d\omega'\frac{{\rm Im}
\Pi^R(\omega',\vecq)}{\omega-\omega'}.
\ee
Using these quantities one can construct an effective theory 
of collective excitations. Their (full, interacting) 
propagators are completely determined 
by the spectral function of the collective excitations
\be 
B(\omega, \vecq) = \frac{-2{\rm Im} \Pi^R(\omega,\vecq)}
 {\left[\omega^2-\vecq^2  -{\rm Re}\Pi^R(\omega,\vecq) \right]^2
+{\rm Im} \Pi^R(\omega,\vecq)^2}.
\ee
The dispersion relation of the collective excitations is 
read-off as $\omega^2 = \vecq^2  +{\rm Re}\Pi^R(\omega,\vecq)$.
The finite life-time effects are described by the width of the 
spectral function, \ie by the function ${\rm Im} \Pi^R(\omega,\vecq)$.

Figure~\ref{fig:reimpi} illustrates the dependence 
of the real and imaginary parts of the pair breaking 
polarization tensor on the transferred energy  
for fixed three-momentum transfer. 
The value of the Landau parameter is  $f_0=-0.5$~\cite{Sedrakian:2006ys}.
The zero temperature gap at $p_F = 0.1 $ fm$^{-1}$ 
is $\Delta (0) = 1$ MeV and  $T_c = \Delta(0)/1.76$. 
The frequencies are normalized 
to the zero-temperature threshold frequency $2\Delta(0)$, 
the momentum transfer to the Fermi momentum.  The real and 
imaginary parts of the polarization tensor scale as $q^2$. Their 
behavior at negative energies follows from their even and 
odd parity with respect to the energy transfer, \ie
${\rm Re}\Pi^R(-\omega) = {\rm Re}\Pi^R(\omega)$ and 
${\rm Im}\Pi^R(-\omega) = -{\rm Im}\Pi^R(\omega)$.
Note that the imaginary parts are identically zero below the
threshold for pair breaking process $2\Delta(T)$.

\begin{figure}[t]
\begin{center}
\vskip -2.0cm
\includegraphics[width=8.0cm,height=10.0cm,angle=-90]{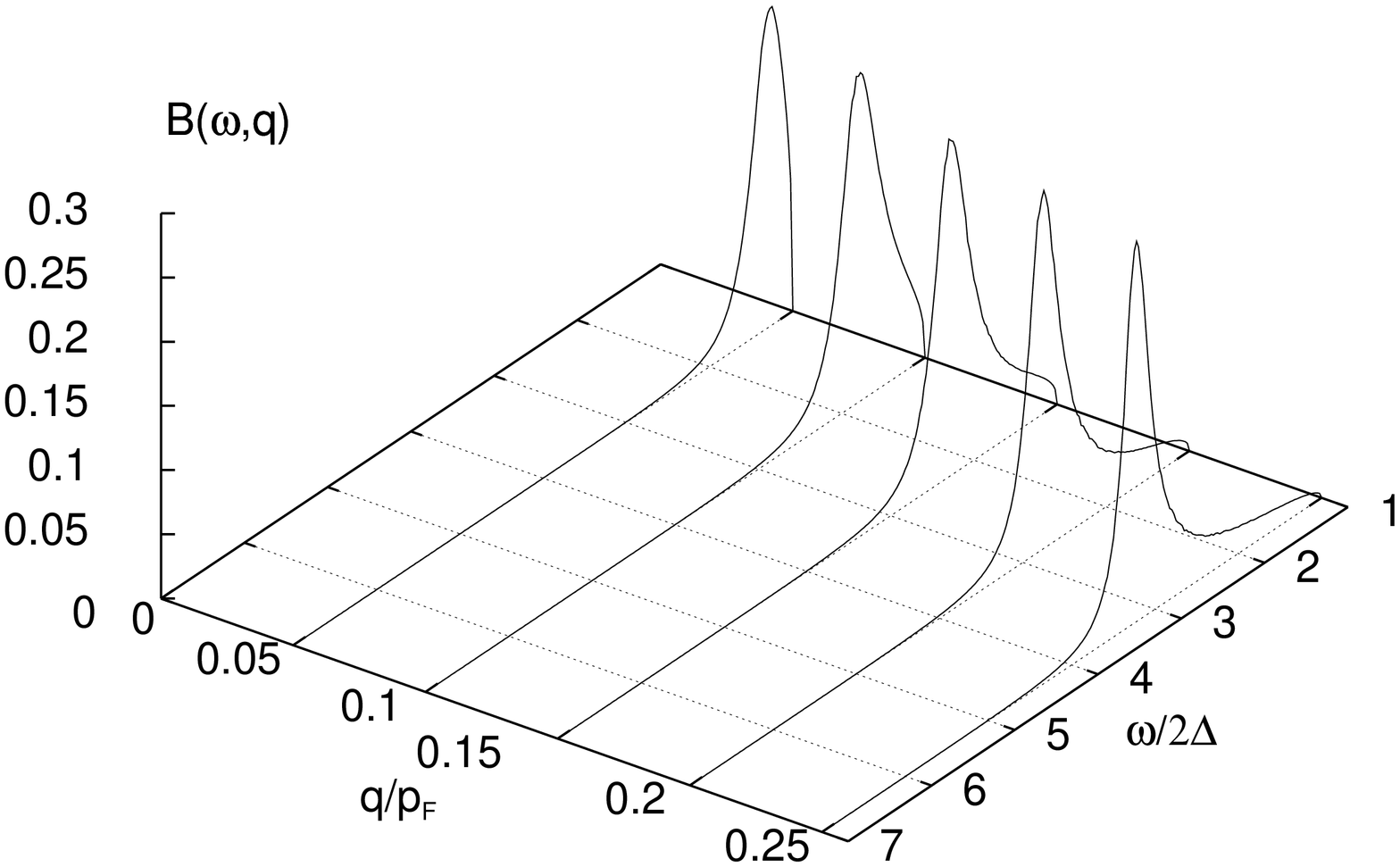}
\vskip -1.0cm
\includegraphics[width=8.0cm,height=10.0cm,angle=-90]{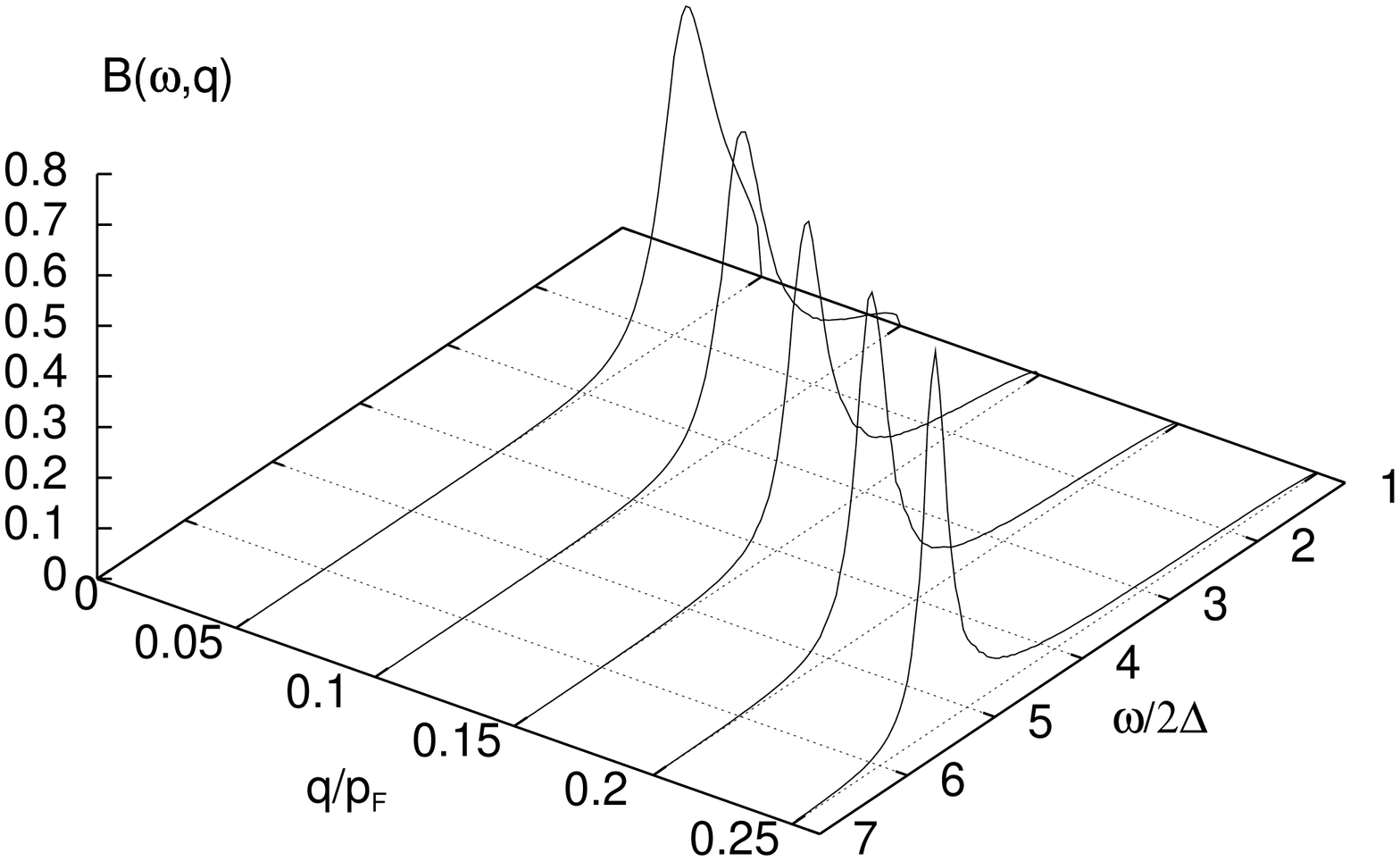}
\end{center}
\caption[]
{The spectral function of density fluctuations at $T= 0.2 T_c$ 
(upper panel) and $T= 0.9 T_c$ (lower panel). The parameters
are as in Fig.~\ref{fig:reimpi}.
}\label{fig:spec_func}
\end{figure}
\begin{figure}[tb]
\begin{center}
\includegraphics[width=\linewidth,height=7.0cm,angle=0]{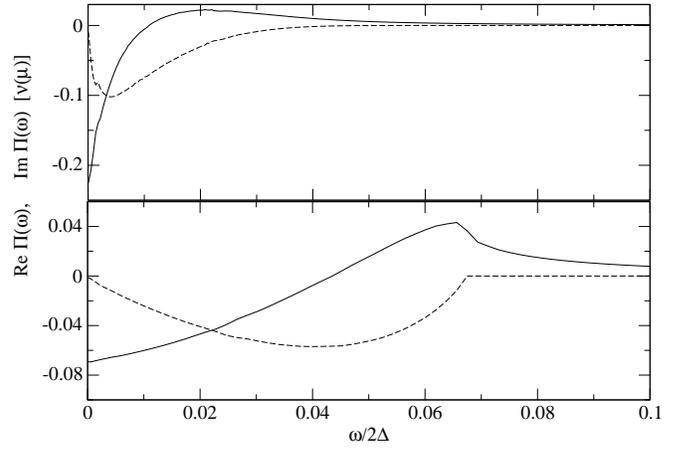}
\end{center}
\caption[]
{Dependence of the real (solid lines) and imaginary (dashed lines)  
parts of the scattering polarization tensor in units of density 
of states on the energy transfer in units of threshold energy 
$2\Delta(0)$ for fixed  momentum transfer $q=0.1 p_F$, 
with $p_F = 0.1$ fm$^{-1}$,  and two temperatures 
 $T = 0.2 T_c$ (upper panel) and $T = 0.9T_c$ (lower panel).
The zero temperature  gap is taken to be $\Delta (0) = 1$ MeV, 
$T_c = \Delta(0)/1.76$. 
}\label{fig:scattering}       
\end{figure}
\begin{figure}[t]
\begin{center}
\vskip -2.0cm
\includegraphics[width=8.0cm,height=10.0cm,angle=-90]{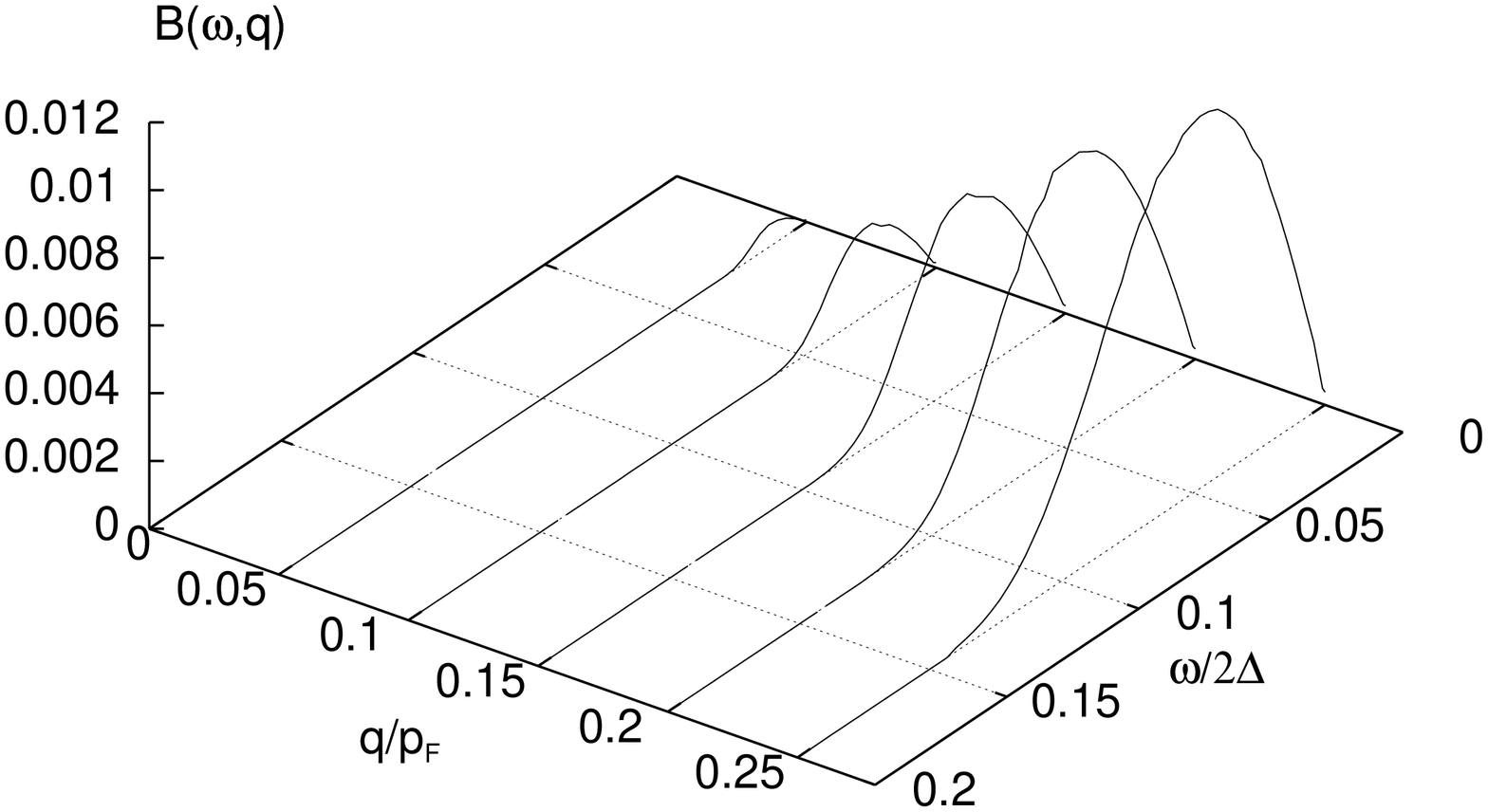}
\vskip -1.0cm
\includegraphics[width=8.0cm,height=10.0cm,angle=-90]{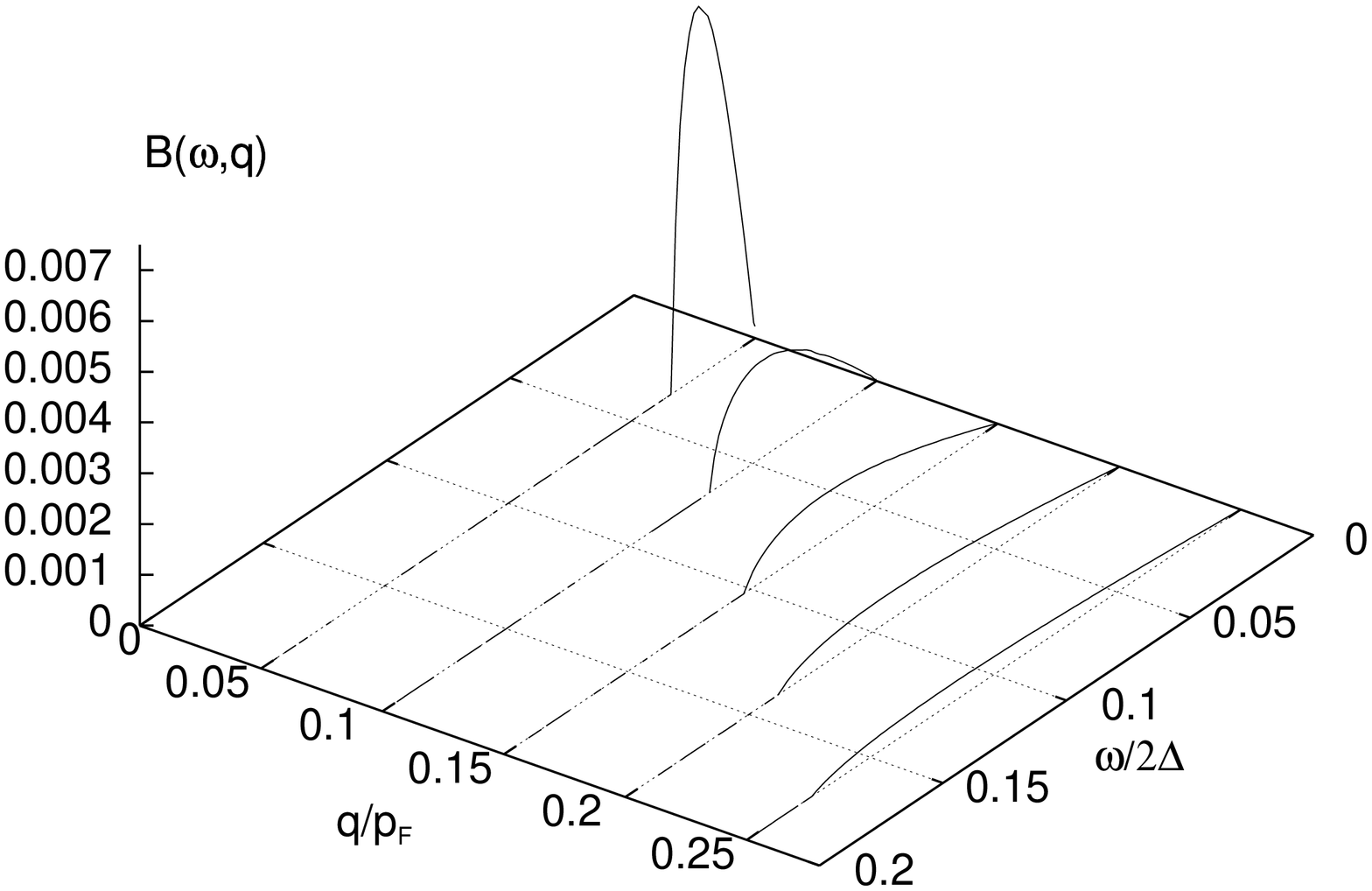}
\end{center}
\caption[]
{The spectral function of density fluctuations at $T= 0.2 T_c$ 
(upper panel) and $T= 0.9 T_c$ (lower panel). The parameters
are as in Fig.~\ref{fig:reimpi}.
}\label{fig:spec_func2}
\end{figure}
Figure~\ref{fig:spec_func} illustrates the spectral functions
of pair breaking density fluctuation on the energy and 
momentum transfers.  The parameters are the same as in 
Fig.~\ref{fig:reimpi}. The form of the spectral function suggests 
that at low temperatures the low-momentum-transfer contribution 
is concentrated near the pair breaking threshold; for larger momentum 
transfers, modes away from the energy threshold become important. 
At higher temperatures ($T\le T_c$) and for any given momentum transfer,
the main contribution to the spectral function comes from higher energy 
modes and the peak values are larger in the latter regime. 

\subsection{Scattering response function: numerical evaluation}
\label{sec:scatter}
The scattering part of the response function is kinematically 
important for the space-like processes and vanishes automatically
in the limit $\vecq\to 0$.  Small momentum expansion of the 
previous section was found inappropriate for the scattering part 
of the polarization function and it was evaluated numerically 
by adapting the method described in Ref.~\cite{Sedrakian:2004qd}. 
On carrying out the angular integrals we are left with
a one-dimensional integral over the energy. As an example, we give 
the expression for the elementary loop
\bea\label{eq:scatt_GG}
&&{\rm Im} \Pi'_{GG}(q)\nu (p_F)^{-1}, 
= - \frac{\pi p_F T}{4q\mu_0}
\int_{-\mu/T}^{\infty}\frac{d\xi_p}{T} \nonumber\\
&&\Biggl[ u_p^2u_{p+q}^2 
\frac{\ep_{p+q}}{\xi_{p+q}}\Bigg|_{x=x_0^+}
\left[f(\ep_p)-f(\ep_{p}+\omega)\right] 
\theta(1-\vert x_0^+\vert)
\nonumber\\
&&-v_p^2v_{p+q}^2 
\frac{\ep_{p+q}}{\xi_{p+q}} \Bigg|_{x=x_0^-}
\left[f(\ep_p)-f(\ep_{p}-\omega)\right]\theta(1-\vert x_0^-\vert)
\Biggr],\nonumber\\
\eea
where $\theta$ is the Heaviside step function,  
$x = (\vecp\cdot \vecq)/\vert\vecp\vert\vert\vecq\vert$, and 
$x_0^{\pm}$ is  the $x$  value satisfying the equation 
$\pm\omega+\ep_p-\ep_{p+q} = 0$ and the prime refers to the 
scattering part. The expressions for the 
scattering parts of the polarization tensors $\Pi'_{XX'}(q)$ 
are similar to  Eq.~(\ref{eq:scatt_GG}), but involve different 
combinations of coherence factors. The real parts are obtained
from the dispersion relation (\ref{KK_relation}).

Figure \ref{fig:scattering} shows the real and imaginary 
parts of the scattering polarization tensor at two temperatures 
and for fixed momentum transfer. Figure \ref{fig:spec_func2} shows the 
spectral function derived from the scattering polarization tensor. It is 
seen that the modes with energies within the breaking threshold
$\omega \le 2\Delta(0)$ are relevant for scattering processes, as opposed 
to the pair breaking case where only the modes with $\omega \ge 2\Delta(0)$
emerge.
This justifies the separation of the modes into two classes. The peak 
values are larger at higher temperatures $T\le T_c$, as is the case 
for the pair breaking response.

\subsection{Unpaired and uncorrelated limits}

\label{sec:unpaired}
Here we wish to obtain the unpaired  ($T>T_c$)
limit of the polarization function (\ref{FULL_PI}).
This amounts to setting the coherence factors in 
Eq.~(\ref{A1})--(\ref{A5}) to their values in the 
normal state
\be
u_p = 1, \quad v_p = 0.
\ee
It follows then from Eq. (\ref{A2bis}) that $\Pi_{FG}= 0 =\Pi_{GF}$
and, therefore,  ${\cal B} = 0$. Setting ${\cal B} = 0$ in 
Eq.~(\ref{FULL_PI}) and noting that ${\cal C} \neq 0$, since
$\Pi_{G^{\dagger}G}\neq 0$, we obtain
\be \label{FULL_PI2}
\Pi_{RPA} (q) = \frac{{\cal A}(q)}{1 -v_{ph}{\cal A}(q)},
\ee
where the function ${\cal A}(q)$ in the unpaired state reduces to
\bea\label{FULL_PI3}
{\cal A}(q) = \Pi_{GG}(q,\Delta=0) 
= g\int\frac{d^3p}{(2\pi)^3}  
\frac{f(\ep_p)-f(\ep_k)}{\omega+\ep_p-\ep_k+i\delta}.\nonumber\\
\eea
Equations (\ref{FULL_PI2}) and (\ref{FULL_PI3}) 
are the standard expressions for the 
polarization tensor of a Fermi liquid in
the random phase approximation (RPA). The free Fermi gas
result follows on setting $v_{ph} = 0$ in Eq. (\ref{FULL_PI2}):
\be
\Pi_{free}(q)  = g\int\frac{d^3p}{(2\pi)^3}
\frac{f(\ep_p)-f(\ep_k)}{\omega+\ep_p-\ep_k+i\delta}.
\ee
Clearly, the pair-braking contribution vanishes as $T\to T_c^{(-)}$. 
The scattering contribution at $T\ge T_c$ reproduces the unpaired
and uncorrelated limits, as it should.

\section{Conclusions}
\label{sec:conclude}

In this article we have carried out several steps in the program 
aimed at understanding the polarization tensor of pair-correlated 
neutron and nuclear matter by (i) constructing a 
low-transfered-momentum approximation to the pair breaking 
polarization tensor and (ii) by numerically evaluating the 
scattering part of the polarization tensor.

Our main result is that the low-momentum expansion of the 
pair breaking polarization tensor starts at quadratic order 
in the ratio of the momentum-transfer to the Fermi momentum. The 
expansion coefficient is universal, \ie depends only the two 
relevant scales: the Fermi energy and the pairing gap. We also 
clarified the structure of the integral equations for the 
vertex functions when the particle-particle and particle-hole
interactions do not coincide and verified explicitly that 
the unpaired and uncorrelated limits are recovered. 

Further steps will require a verification of the convergence 
of the series by a comparison of our analytical results with 
an exact numerical evaluation of the response functions.
The rate of series convergence can be checked by computing the 
next-to-leading-order contribution.
Further refinements could include
finite range interactions, tensor forces, and so on.
The new numerical and analytical methods, discussed in this article,
could be useful in the studies of the response functions of 
pair-correlated fermionic systems in general.

The main implication of our study concerns the neutrino emissivity 
via the vector current pair breaking bremsstrahlung, which are 
phenomenologically important in the physics of neutron star 
cooling and superbursts  in accreting neutron 
stars~\cite{Brown:2009kw,Page:2009fu,Gupta:2006fd}. Our results 
suggest that the vector current contribution to the process
\be\label{PB}
\{nn\} \to \{nn\}+\nu+\anu ,
\ee
where $\{nn\}$ refers to the pair correlated state, 
$\nu$ and $\anu$ to the neutrino and antineutrino,
is suppressed to a lesser degree, than suggested 
in the recent literature~\cite{Leinson:2006gf,Kolomeitsev:2008mc}.
However, further studies indicated above are needed to draw 
a final conclusion on the relative importance of this vector 
current neutrino emission process (\ref{PB}).

\acknowledgments

We thank T. Brauner, X.-G. Huang, and D. H. Rischke   
for discussions. The work of J. K. was supported by 
the Deutsche Forschungsgemeinschaft
(Grant SE 1836/1-1).

\appendix
\section{Expanding the functions ${\cal A}(q)$, 
${\cal B}(q)$, and ${\cal C}(q)$ }
\label{appendix_A}
We write each of these functions as a product of an 
appropriate coherence factor and corresponding statistical factor
\bea\label{A2}
{\cal A}(q)  &=&  \int\frac{d^3p}{(2\pi)^3}
a(\vecp,\vecq,\omega)L(\vecp,\vecq,\omega),\\
\label{B2_2}
{\cal B}(q) &=&\Delta\omega \int\frac{d^3p}{(2\pi)^3}
\frac{1}{\ep_p} L(\vecp,\vecq,\omega),\\
\label{C2}
{\cal C}(q) &=& -\int\frac{d^3p}{(2\pi)^3}  c(\vecp,\vecq,\omega)
L(\vecp,\vecq,\omega)\nonumber\\
&-& \int\frac{d^3p}{(2\pi)^3} \frac{1}{2\ep_p}\left[1-2f(\ep_p)\right],
\eea
where
\bea
a(\vecp,\vecq,\omega) &=& \frac{\ep_p+\ep_{p+q}}{2}
\left(1-\frac{\xi_p\xi_{p+q}}{\ep_p\ep_{p+q}}+ \frac{\Delta^2}{\ep_p\ep_{p+q}}
\right),\nonumber\\
\\
c(\vecp,\vecq,\omega) &=&
\frac{\ep_p+\ep_{p+q}}{2}\left(1+\frac{\xi_p\xi_{p+q}}{\ep_p\ep_{p+q}}
+\frac{\Delta^2}{\ep_p\ep_{p+q}}\right),\nonumber\\
\\
L(\vecp,\vecq,\omega)
&=& \frac{1-f(\ep_p)-f(\ep_{p+q})}{(\omega+i\delta)^2-(\ep_p+\ep_{p+q})^2}.
\eea
Next we expand these functions in powers of small parameter $y$ to 
order $O(y^2)$
\bea
\label{a_exp}
a(\vecp,\vecq,\omega) &=& \frac{2\Delta^2}{\ep_p}
-\frac{2\Delta^2\,\mu\,\xi_p}{\ep_p^3} ~x~y\nonumber\\
&+&\frac{\Delta^2\,\mu\,\xi_p}{\ep_p^5}(6\mu\xi_px^2 -\ep_p^2 )~y^2,\\
\label{c_exp}
c(\vecp,\vecq,\omega) &=& 2\ep_p + \frac{2\mu\xi_p}{\ep_p} ~x~y
+\frac{\mu\,\xi_p}{\ep_p}~y^2,\\
\label{L_exp}
L(\vecp,\vecq,\omega) &=& L_0 + L_1 x y + L_2y^2,
\eea
where the coefficients of the expansion (\ref{L_exp}) can be 
written using the shorthand expressions $D\equiv \omega^2 - 4\ep_p^2$
and $\zeta \equiv \ep_p/2T$ as
\bea
\label{L0}
L_0 &=& \frac{\,\tanh \,\zeta}{D+i\delta},
\eea
\bea
\label{L1}
L_1 &=& 8\,\mu\,{\xi_p}\,\frac{\tanh\,\zeta}
   {{\left( D+i\delta \right) }^2} +
 \frac{\mu\,{\xi_p}}{2{\ep_p}\,T}  \frac{{\rm sech}^2\,\zeta}{(D+i\delta)},
\eea
\bea\label{L2}
L_2 &=& \frac{4\,{\mu}^2\,x^2\,{{\xi_p}}^2\,
    {\rm sech}^2\,\zeta}
   {{\ep_p}\,T\,{\left( D+i\delta \right) }^2}
+   \frac{{\mu}^2\,x^2\,{\rm sech}^2\,\zeta}
{2\,{\ep_p}\,T\,\left(D+i\delta \right) }
\nonumber\\
&+& 
\frac{\mu\,{\xi_p}{\rm sech}^2\,\zeta}{4\,{\ep_p}\,T\,\left( D+i\delta\right)}
-   \frac{{\mu}^2\,x^2\,{{\xi_p}}^2\,{\rm sech}^2\,\zeta}
    {2\,{{\ep_p}}^3\,T\,\left( D+i\delta \right) } \nonumber\\
&+&
  \frac{64\,{\mu}^2\,x^2\,{{\xi_p}}^2\,
     \tanh\,\zeta}{{\left( D+i\delta \right) }^3} +
  \frac{8\,{\mu}^2\,x^2\,\tanh\,\zeta}
   {{\left( D+i\delta \right) }^2} \nonumber\\
&+&
  \frac{4\,\mu\,{\xi_p}\,\tanh \zeta}
   {{\left(D+i\delta \right) }^2} 
-\frac{4\,{\mu}^2\,x^2\,{{\xi_p}}^2\,
     \tanh\zeta}{{{\ep_p}}^2\,
     {\left( D+i\delta \right) }^2}\nonumber\\
&-& \frac{{\mu}^2\,x^2\,{{\xi_p}}^2\,
\tanh \zeta\, {\rm sech}^2\zeta}{2\,{{\ep_p}}^2\,T^2\, 
\left( D+i\delta \right) } ,
\eea
where ${\rm sech}^2\left(x\right) \equiv
1 - \tanh \left(x\right)^2$. On substituting  
Eqs.~(\ref{a_exp}) and (\ref{L_exp}) 
in Eq.~(\ref{A2}) we obtain
\bea
{\cal A}(q)  &=&  \int\frac{d^3p}{(2\pi)^3}
\Biggl[\frac{2\Delta^2}{\ep_p}L_0
+\frac{2\Delta^2}{\ep_p}L_2 y^2\nonumber\\
&-&\frac{2\Delta^2\,\mu\,\xi_p}{\ep_p^3}~L_1 ~x^2~y^2
+\frac{\Delta^2\,\mu\,\xi_p}{\ep_p^5}(6\mu\xi_px^2 -\ep_p^2 )L_0~y^2
\Biggr]
\nonumber\\
&=&{\cal A}_0(q)  + {\cal A}_2(q)y^2+O(y^4).
\eea
The terms that are odd in $x$ drop out on integration 
in symmetrical limits. Similarly,
\bea
\label{C2_2}
{\cal C}(q) &=& -\int\frac{d^3p}{(2\pi)^3}
\Biggl[2\ep_p L_0 + \frac{2\mu\xi_p}{\ep_p}L_1 ~x^2~y^2
+\frac{\mu\,\xi_p}{\ep_p}~L_0~y^2 \nonumber\\
&+& 2\ep_p L_2y^2\Biggr]- \int\frac{d^3p}{(2\pi)^3} 
\frac{1}{2\ep_p}\left[1-2f(\ep_p)\right].
\eea
We further use the relation  
\bea
&&-\int\frac{d^3p}{(2\pi)^3}  
2\ep_p L_0 - \int\frac{d^3p}{(2\pi)^3} 
\frac{1}{2\ep_p}\left[1-2f(\ep_p)\right]\nonumber\\
&&\hspace{3.cm}=-\frac{\omega^2}{2}\int\frac{d^3p}{(2\pi)^3}
 \frac{1}{\ep_p}L_0\,
\eea
to write 
\bea
\label{C2_4}
{\cal C}(q)
 &=& -\frac{\omega^2}{2}\int\frac{d^3p}{(2\pi)^3}
 \frac{1}{\ep_p}L_0 -\int\frac{d^3p}{(2\pi)^3}
\Biggl[ \frac{2\mu\xi_p}{\ep_p}L_1 ~x^2~y^2\nonumber\\
&+&\frac{\mu\,\xi_p}{\ep_p}~L_0~y^2 + 2\ep_p L_2y^2
\Biggr]={\cal C}_0(q)  + {\cal C}_2(q)y^2.\nonumber\\
\eea
Finally,
\bea
{\cal B}(q) 
&= &
\Delta\omega \int\frac{d^3p}{(2\pi)^3}
\frac{1}{\ep_p} (L_0 + L_2~y^2) = {\cal B}_0(q)  + {\cal B}_2(q)y^2,
\nonumber\\
\eea
and ${{\cal A}_0}/{{\cal C}_0} = -{4\Delta^2}/{\omega^2},$ and 
${{\cal B}_0}/{{\cal C}_0} =-{2\Delta}/{\omega}$.

\section{Evaluating phase-space integrals}
\label{appendix_B}
Here we evaluate the integrals (\ref{int_I0})--(\ref{int_I2}), 
which are given as 
\bea\label{I0}
I_0 &=& \nu(p_F)^{-1}\mu\,\int\frac{d^3p}{(2\pi)^3}
\frac{\xi_p}{\ep_p^3}\left[\frac{4\ep_p^2}{\omega^2}-1
+\frac{6\mu\xi_p}{\ep_p^2}x^2\right]L_0,\nonumber\\\\
\label{I1}
I_1 &=& 2\nu(p_F)^{-1}\mu\int\frac{d^3p}{(2\pi)^3}
\frac{\xi_p}{\ep_p^3}
\left[\frac{4\ep_p^2}{\omega^2} -1\right]~L_1 x^2,\\
\label{I2}
I_2 &=& 2\nu(p_F)^{-1}\int\frac{d^3p}{(2\pi)^3}
 \frac{1}{\ep_p}\left[\frac{4\ep_p^2}{\omega^2}-1\right]L_2(x).
\eea
After carrying out the angular integrals we obtain
\bea
I_0 &=& \nu(p_F)^{-1}\mu\,\int\frac{dpp^2}{2\pi^2}
\frac{\xi_p}{\ep_p^3}\left[\frac{4\ep_p^2}{\omega^2}-1
+\frac{2\mu\xi_p}{\ep_p^2}\right]L_0,\nonumber\\\\
I_1 &=& \frac{2}{3}\nu(p_F)^{-1}\mu\int\frac{dpp^2}{2\pi^2}
\frac{\xi_p}{\ep_p^3}
\left[\frac{4\ep_p^2}{\omega^2} -1\right]~L_1 ,\\
I_2 &=& 2\nu(p_F)^{-1}\int\frac{dpp^2}{2\pi^2}
 \frac{1}{\ep_p}\left[\frac{4\ep_p^2}{\omega^2}-1\right]\langle L_2(x)\rangle ,
\eea
where $\langle L_2\rangle $ is the angle integrated loop $L_2$,
which is obtain from (\ref{L2}) via the substitution $x^2 = 1/3$.
Using the transformation $
dp p^2 = m^*p_F d(p^2/2m^*) = m^*p_F d\xi_p,$ and the relation
$\ep_pd\ep_p = \xi_pd\xi_p$ we  obtain
\bea
I_0 &=& \mu\,\int_{\Delta}^{\infty}{d\ep_p}
\frac{1}{\ep_p^2}\left[\frac{4\ep_p^2}{\omega^2}-1
+\frac{2\mu\xi_p}{\ep_p^2}\right]L_0,\\
I_1 &=& \frac{2\mu}{3}\int_{\Delta}^{\infty}{d\ep_p}
\frac{1}{\ep_p^2}
\left[\frac{4\ep_p^2}{\omega^2} -1\right]~L_1,\\
I_2 &=& 2\int_{\Delta}^{\infty}{d\ep_p}
 \frac{1}{\xi_p}\left[\frac{4\ep_p^2}{\omega^2}-1\right]\langle L_2(x)\rangle.
\eea
To make further progress we need to separate the real and imaginary
parts of the integrals. We shall first compute the imaginary parts.
They are extracted with the help of the identity
\be
\frac{1}{(D+i\delta)^{n+1}} = \frac{P}{D^{n+1}}
-i\pi \frac{(-1)^n}{n!}\delta^{(n)}(D),
\ee
where $P$ denotes the principal value and $\delta^{(n)}(D)$ is
the $n$-th derivative of the delta function.

For positive $\omega$ the integral  (\ref{I0}) 
gives~\footnote{We use the identity $\delta(x^2-a^2) =\vert
2a\vert^{-1}[\delta(x+a)+\delta(x-a)]$.}
\bea
\label{eq:int0final}
{\rm Im}I_0 &=& -\pi \mu\,\int_{\Delta}^{\infty}{d\ep_p}
\frac{1}{\ep_p^2}\Biggl[\frac{4\ep_p^2}{\omega^2}-1\nonumber\\
&+&\frac{2\mu\xi_p}{\ep_p^2}\Biggr]
\tanh \left(\frac{\ep_p}{2\,T}\right)
\delta(\omega^2-4\ep_p^2)\nonumber\\
 &=&  -8\pi \mu^2\,\frac{\sqrt{\omega^2-4\Delta^2} 
~{\rm sgn}(\omega)}{\omega^5}
\tanh \left(\frac{\omega}{4\,T}\right)\,
\theta(\omega-2\Delta).\nonumber\\
\eea
Consider the integral (\ref{I1}). First, note that the 
term $\propto D^{-1}$  vanishes, since the delta function 
enforces $\omega^2 = 4\ep_p^2$. After dropping this term 
we are left with the integral
\bea
\label{I1a}
{\rm Im}I_1 &=& \frac{16\pi\mu^2}{3}
\int_{\Delta}^{\infty}d\ep_p \frac{1}{\ep_p^2}
\left(\frac{4\ep_p^2}{\omega^2}-1\right)\sqrt{\ep_p^2-\Delta^2}
\nonumber\\&&
\tanh\left(\frac{\ep_p}{2T}\right)\delta^{(1)}(\omega^2-4\ep_p^2).
\eea
This and similar integrals, which contain derivatives of the delta 
function, are computed via the formula
\be\label{rule} 
\int f(x)\delta^{n}(x-a)dx = (-)^nf^{(n)}(a).
\ee
The result of integration is
\be
\label{eq:int1final}
{\rm Im}I_1 = -\frac{8\pi\mu^2}{3}
\frac{\sqrt{\omega^2-4\Delta^2}~{\rm sgn}(\omega)
}{\omega^5}\tanh\left(\frac{\omega}{4T}\right)\,
\theta(\omega-2\Delta)\,.
\ee
In the integral (\ref{I2}) we again omit terms $\propto D^{-1}$,
since their prefactors are zero after integration. 
After inserting $x^2=1/3$ in the remainder we obtain
\begin{widetext}
\bea
{\rm Im}I_2 &=&
\frac{8\pi}{3}\int_{\Delta}^{\infty}d\ep_p\,
  \left[\frac{4\ep_p^2}{\omega^2}-1\right]
  \Bigg\{\frac{\mu^2\sqrt{\ep_p^2-\Delta^2}}{\ep_p\,T}{\rm sech}^2
\left(\frac{\ep_p}{2T}\right)
         +\frac{2}{\sqrt{\ep_p^2-\Delta^2}}
\mu^2\tanh\left(\frac{\ep_p}{2T}\right)\nonumber\\
&&\hspace{20mm}         
         +3\mu\tanh\left(\frac{\ep_p}{2T}\right)
         -\frac{\mu^2\sqrt{\ep_p^2-\Delta^2}}
{\ep_p^2}\tanh\left(\frac{\ep_p}{2T}\right)
  \Bigg\}\,\delta^{(1)}(\omega^2-4\ep_p^2)\nonumber\\
&& -\,\frac{64\pi\mu^2}{3}\int_{\Delta}^{\infty}d\ep_p\,
  \left[\frac{4\ep_p^2}{\omega^2}-1\right]
  \sqrt{\ep_p^2-\Delta^2}\tanh\left(\frac{\ep_p}{2T}\right)
  \delta^{(2)}(\omega^2-4\ep_p^2)\,.
\eea
Applying Eq.~(\ref{rule}) we obtain 
\bea\label{eq:int2final}
{\rm Im}I_2 
&=&-\frac{\pi\mu ~{\rm sgn}(\omega)}
{3T\omega^5\sqrt{\omega^2-4\Delta^2}}
\,{\rm sech}^2\left(\frac{\omega}{4T}\right)
  \Bigg\{T\,[24\Delta^2\mu+\omega^2\,(2\mu+3\sqrt{\omega^2-4\Delta^2})]
  \sinh\left(\frac{\omega}{2T}\right)
 +4\mu\omega\,(\omega^2-4\Delta^2)
\Bigg\}\,\theta(\omega-2\Delta)\,.\nonumber\\
\eea
Finally, adding the three integrals 
(\ref{eq:int0final}), (\ref{eq:int1final}),
and (\ref{eq:int2final}) we obtain Eq.~(\ref{calP_final})
of the main text.
\end{widetext}

\end{document}